\documentclass[journal]{IEEEtran}
\usepackage{graphicx}
\usepackage{url}
\usepackage{commath}
\usepackage{booktabs} 
\usepackage{algorithm2e}
\usepackage{cite}
\usepackage{multicol}
\usepackage{amsmath}

\ifCLASSINFOpdf
\else
\fi
\begin{document}
%
\title{A Secure Back-up and Restore for Resource-Constrained IoT based on Nanotechnology}
%
%
%
\author{Mesbah~Uddin,
        Md~Badruddoja~Majumder,
        Md~Sakib~Hasan,
        and~Garrett~S.~Rose
\thanks{This material is based upon work supported by the Air Force Office of Scientific Research under award number FA9550-16-1-0301. Any opinions, finding, and conclusions or recommendations expressed in this material are those of the authors and do not necessarily reflect the views of the United States Air Force.}
\thanks{M. Uddin, M. B. Majumder, and Garrett S. Rose are with the Department of Electrical Engineering and Computer Science, University of Tennessee, Knoxville, TN, 37996 USA (e-mail: [muddin6,mmajumde,garose)@utk.edu].}
\thanks{M. S. Hasan is with the Department of Electrical Engineering in the University of Mississippi, Oxford, MS 38677 (email: mhasan5@olemiss.edu).}

}

%
%

\markboth{
}%
{Uddin \MakeLowercase{\textit{et al.}}: }
%



\maketitle

\begin{abstract}
With the emergence of IoT (Internet of things), huge amounts of sensitive data are being processed and transmitted everyday in edge devices with little to no security. Due to their aggressive power management schemes, it is a common and necessary technique to make a back-up of their program states and other necessary data in a non-volatile memory (NVM) before going to sleep or low power mode. However, this memory is often left unprotected as adding robust security measures tends to be expensive for these resource constrained systems. In this paper, we propose a lightweight security system for NVM during low power mode. This security architecture uses the memristor, an emerging nanoscale device which is used to build hardware security primitives like PUF (physical unclonable function) based encryption-decryption, true random number generators (TRNG), and memory integrity checking. A reliability enhancement technique for this PUF is also proposed which shows how this system would work even with less-than-100\% reliable PUF responses. Together, with all these techniques, we have established a dual layer security protocol (data encryption+integrity check) which provides reasonable security to an embedded processor while being very lightweight in terms of area, power, and computation time. A complete system design is demonstrated with 65$n$m CMOS and emerging memristive technology. With this, we have provided a detailed and accurate estimation of resource overhead. Analysis of the security of the whole system is also provided.

\end{abstract}

\begin{IEEEkeywords}
IoT security, hardware security, IoT, embedded system security, physical unclonable function, PUF, emerging nanotechnology, memristor, RRAM.
\end{IEEEkeywords}

%
\IEEEpeerreviewmaketitle

\section{Introduction}


\IEEEPARstart{B}{illions} of smart IoT devices are introduced into our lives every year \cite{IoT-device-number:2015}. In coming years, this number will only increase and with it the amount of sensitive information such systems carry. However, most of these devices are equipped with very little to no security. Therefore, it is not too difficult to tap into these devices and gather sensitive information. Data can be leaked from hardware architectures as well as the apps or software frameworks that these devices use, thereby compromising safety and privacy \cite{Sikder_IoT_privacy_survey:2018,Acar_IoT_peekaboo:2018}. While there are works that target at how to restrict the use of such data and mitigate data leaks \cite{Oconnor_IoT_homesnitch:2019,Babun_IoT_privacy:2019} to maintain privacy, we are mainly concerned about the standalone data and memory security of such devices in this particular work. There is a growing need to employ security measures in these devices. However, many IoT devices are usually resource constrained, battery operated or even batteryless and small in size. They usually employ very aggressive power saving techniques and often go into ultra low power or sleep mode to save energy. It is often necessary to back-up their states, registers and other information into a non-volatile memory (NVM) before going into sleep so they can resume their operation quickly when needed. Moreover, for batteryless energy-harvesting devices, power failure may occur frequently such that it is beneficial to back-up often \cite{Ma:Energy_harvesting_architecture_explore:2015}. NVM can retain their states with zero standby power making it the ideal choice for storing the back-up data. However, during a power failure or when the device is in energy-saving mode, data left on the non-volatile memory may be unprotected and might reveal sensitive information to an adversary. In this work, we have designed a system to provide lightweight security for NVM.

Embedded systems require lightweight cryptographic solution and several such cryptosystems are discussed in \cite{Manifavas_Lightweight_Crypto:2013}. However, even these lightweight solutions would incur an overhead that can be considered too large for resource constrained systems such as IoT. As we have discussed later, our proposed security solution would require both encryption-decryption and hash or tag generation mechanism. Combining traditional cryptographic algorithms like AES \cite{Amir_AES_hardware:2011}, PRESENT \cite{Bogdanov_PRESET:2007}, SIMON and hash functions like SHA-2, SHA-3], MD5 \cite{Manifavas_Lightweight_Crypto:2013} etc. would present a large overhead to these small systems. Moreover, we are also motivated by the notion of providing device-specific unique security solution to each such system. Considering all these, we propose a lightweight encryption scheme using a PUF as the key generator to secure the non-volatile memory of an embedded processor. 

A PUF is a circuit that can generate a unique signature or key specific to a particular implementation of an integrated circuit (IC) \cite{Suh:2005}. Due to uncontrollable manufacturing process variation, there are some differences across different IC implementations or even among different physical locations within an IC. In this work, we use a memristor based PUF which takes a very small area, has smaller delay, and also low power consumption. The memristor (memory+resistor) \cite{Strukov:2008} is an emerging nanoscale device that exhibits a large process variation, useful for implementing PUFs that are very small in size \cite{Mesbah-JETC:2017}. We also consider memristors in the design for the NVM of our demonstrated system in this work.

As PUF responses depend on tiny process variations, reliability is a major concern which limits their usage in key generation applications. Efficient error correction methods are, therefore, essential in PUF based security applications. But existing robust error correction techniques can be computationally expensive as well as resource heavy, thus making them unsuitable for embedded systems. In this work, we instead propose an reliability enhancement technique to improve the reliability of a memristor based PUF and also present a security protocol that relaxes the required level of reliability of a PUF response. After the reliability enhancement block, the PUF response is used as the key to encrypt the back-up data. The same key can be used for fast decryption as well. The idea is to generate different keys from the PUF by applying different challenges each time a back-up operation is needed.


To verify that the stored data has not been altered due to a data error or maliciously by an adversary, we have also used an integrity checking protocol during system wake-up, thereby providing a second layer of security. The integrity checking protocol considered here has in-memory tag generation capability leveraging sneak path currents in a memristive (or any resistive) memory \cite{Majumder_IntegrityCheck:2018}.
After performing a back-up of the program state and other necessary data, a tag is generated from the memory in order to verify data integrity when the system wakes up. This integrity checking system for memristive memory is a good fit for IoT systems as it provides data integrity with significantly lower overhead. The PUF also needs a random challenge on each backup operation. We have decided to use a memristor based TRNG \cite{Mesbah_TRNG:2019} to generate that random challenge vector. 

Among the contributions of this work are:
\begin{itemize}
    \item proposition of a novel hardware based security protocol for embedded system or IoT security,
    \item providing a complete security solution using emerging nanotechnology, specifically memristive technology,
    \item design of a lightweight reliability enhancement technique for memristor based PUFs while relaxing the reliability requirement for unique key generation,
    \item complete transistor-level implementation of the whole system for an accurate resource requirement estimation and overhead analysis.
\end{itemize}

The paper is organized as follows: section \ref{sec:backgnd} first gives a quick introduction to memristors and PUFs. Section \ref{sec:proposed-protocol} describes the key idea of this work and section \ref{sec:protocol} presents the proposed security protocol. Section \ref{sec:reli_enhance} presents the theory and analyses for reliability enhancement technique used in this work. Section \ref{sec:system-design} describes different components of our proposed security architecture. Section \ref{sec:probable_attacks} presents a list of probable attack scenarios where section \ref{sec:security_properties} provides a security analyses of individual components of our design. Section \ref{sec:security_results} provides a detailed security analysis against the attacks list in section \ref{sec:probable_attacks} while section \ref{sec:performance} discusses the resource overhead. Finally, future prospects and concluding remarks are provided in sections \ref{sec:future} and \ref{sec:conc}, respectively.


\section{Background}
\label{sec:backgnd}
\subsection{Memristor}
The memristor, also known as RRAM or ReRAM (resistive random access memory), is one of the most promising emerging nanoscale devices of the last few years \cite{Strukov:2008}. A memristor is a two terminal electrical device whose instantaneous resistance is dependent on its previous history. In a bipolar memristor, resistance can be switched between a high resistance state (HRS) and low resistance state (LRS) by applying an appropriate bias voltage. Switching also requires that the bias voltage be applied above a particular threshold level for at least some minimum amount of time called the switching time. As an emerging and nanoscale device, memristors exhibit large variability in their electrical characteristics which makes them suitable for hardware security applications like PUFs. 
They are very small in size, effectively fitting between cross-points of on-chip interconnect wires (and thus effectively taking zero area), CMOS-compatible, non-volatile in nature, fast ($\sim$ $n$s), have good retention time and endurance \cite{TaOx_Yang:2010,TiO_Medeiros:2011}. Memristors are being considered for a host of different applications in recent years including hardware security \cite{Rose:2013a,Jiang:2017_Memristor_TRNG}, in-memory computation \cite{Hamdioui:2015_memristor_in_memory_compute}, and persistent memories (NVM) \cite{Xu_RRAM_memory:2011,Zidan_memristor_future:2018}.


\begin{figure}
\centering
\hbox{\hspace{-1.3em}
\includegraphics[width=3.8in]{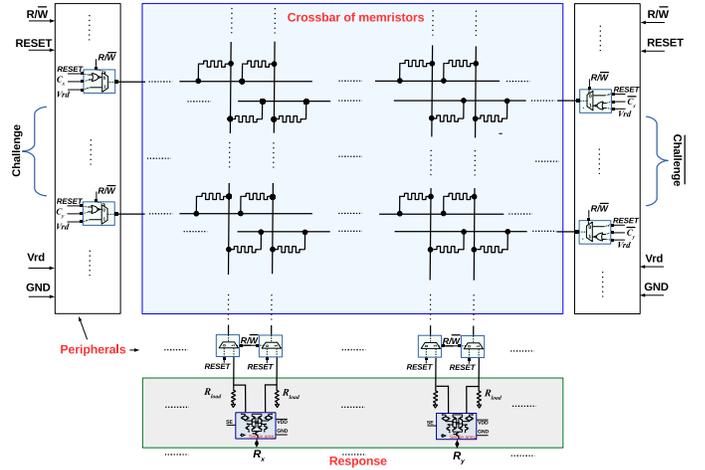}}
\caption{A schematic of the XbarPUF showing memristor crossbar and the peripheral circuitry \cite{Mesbah_SA:2018}}
\label{fig:xbarpuf}
\end{figure}

\subsection{PUF}
Physical unclonable functions (PUF) are promising security primitives useful for a number of security applications, including mitigation of integrated circuit (IC) piracy, cloning, and counterfeit \cite{Suh:2007, paral2011reliable}. A PUF is any circuit that can produce a variety of unique outputs when implemented on different devices corresponding to the same given input. 
The input and output of a PUF are also known as the challenge and response, respectively. PUFs implemented on different chips provide unique challenge-response combinations which can be used as the authenticating signatures for that chip. The PUF concept usually relies on exploiting random and uncontrollable physical entropy sources in a device that make the response unique over various implementations \cite{Suh:2007}. A PUF that has a large number of challenge-response pairs (CRPs) is called a strong PUF such as arbiter PUF (APUF) \cite{Suh:2007} and ring-oscillator PUF \cite{Suh:2005}.

A memristive crossbar PUF or XbarPUF is a strong PUF implemented by a crossbar array of memristors where the switching variability among memristors is harnessed to generate PUF responses. Originally proposed in \cite{Rose:2015}, it was shown to be very light-weight in terms of both area and energy. An improved version of it was presented in \cite{Mesbah_SA:2018}. This is the PUF that we'll be using in this work. An schematic diagram of an XbarPUF is shown in Fig. \ref{fig:xbarpuf}.

\section{Proposed Security Solution for Back-up and Restore in NVM}
\label{sec:proposed-protocol}


\begin{figure}
    \centering
    \hbox{\hspace{-.7em}
    \includegraphics[width=3.5in]{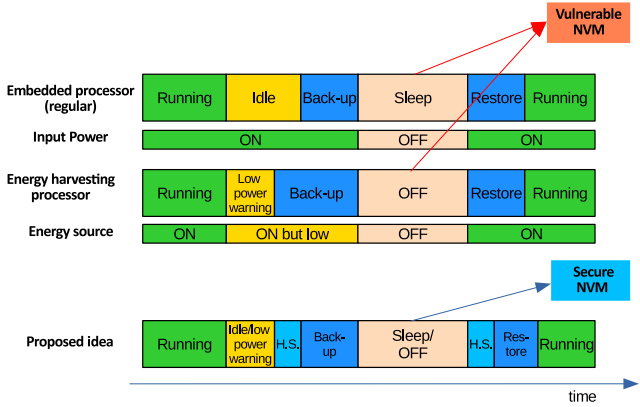}}
    \caption{An embedded processor (either regular or energy harvesting) backs-up data in NVM for data-forwarding but the NVM is left unprotected when there is no power. Proposed idea is to secure this NVM using hardware security.}
    \label{fig:embedded-processor-back-up}
\end{figure}

\subsection{Security vulnerability of IoT/embedded processor}
The goal of this work is to provide security for the unprotected NVM of an IoT edge device after the system processor goes into an ultra low power mode or has a power failure for the case of an energy-harvesting device \cite{Kaisheng_energy_harvest_NVM:2015}. In both scenarios, the NVM is cut off from power and is left unprotected and thus vulnerable to be maliciously read by an adversary to give away sensitive information. This is illustrated in Fig. \ref{fig:embedded-processor-back-up}. To prevent that from happening, we are proposing to secure that NVM by doing a secure back-up before powering down and a secure restore after waking up.

\begin{figure}
    \centering
    \hbox{\hspace{3em}
    \includegraphics[width=2.4in]{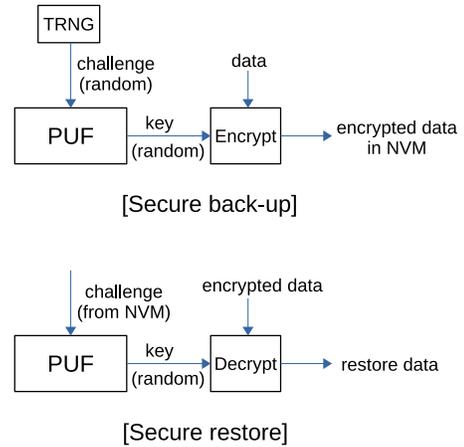}}
    \caption{A random challenge is applied to a PUF to generate a cryptographic key for secure back-up and restoration of data in an embedded processor. A new unique key is generated each time, thus effectively implementing an OTP.}
    \label{fig:puf-based-backup}
\end{figure}

The general idea is to apply a random challenge (e.g. generated from a TRNG) to an on-chip PUF to generate a secure cryptographic key which can be used to encrypt the data before back-up. During wake-up, the same key can be generated again by applying the same challenge to that PUF and data can be decrypted again before restore. This is shown in Fig. \ref{fig:puf-based-backup}. If the challenge applied to a strong PUF is random, the response is random as well and also unique across different devices, thereby providing device-specific security.

In this work, we propose to encrypt the back-up data using a cryptographic key generated by the XbarPUF. As a strong PUF, an XbarPUF can generate a large number of responses to be used as keys. We can generate a unique random key each time a backup operation is needed from this strong PUF and thus effectively implementing an one time pad (OTP) \cite{Roarke_OTP:2013}. Since responses from a PUF might have some unreliable bits, these bits are eliminated during run-time and only reliable bits are used in the final key. Then the data is encrypted with this key and stored in an NVM. A tag is also generated from the stored data for integrity checking purposes. Both this tag and PUF challenge are also stored in an NVM (same as data or a separate secure memory). When power comes back on, a new tag is first generated and checked with the existing tag. If they both match with each other, then the same stored challenge is applied on the XbarPUF again. Similar to back-up operation, unreliable PUF response bits are discarded after reliability enhancement, and a new key (which should be equal to the key generated during backup) is generated. Using this key, the encrypted data from NVM are decrypted and then restored back into their respective registers or other memory location. The whole process is illustrated in Fig. \ref{fig:puf-based-backup}.

Since we are targeting IoT device with which would have resource limitation, instead of using a single large PUF, we have instead proposed to use several smaller PUFs in a time-multiplexed manner. This would increase the system delay, with a reduction in power consumption. A trade-off can be made depending of the resource of a particular device. Besides, if the data to be saved is larger in size, several successive challenges could be applied to and then multiple responses could be appended together for a key as large as this data. More details about this system implementation are also discussed later in this work.


\subsection{System assumptions}
The basic assumptions for implementing such a system are as follows:

\subsubsection{Random key} By definition, the response of a strong PUF is random and unpredictable. The metrics to evaluate a PUF, like bit-aliasing, uniformity etc. are calculated from Monte Carlo simulations to show how random the responses from the PUF that we have used.

\subsubsection{Unique key} PUF responses vary from one implementation to another even for the same challenge, thereby generating device-specific unique key. Thus, even if an adversary can gain access to one chip, he/she won't be able to break the security of another such chip. Uniqueness metric of a PUF represents this quality.

\subsubsection{Key length} The key length should be equal to data length for backup. This is only possible for small embedded system where the amount of backup (only program states and some registers) is small. For slightly larger systems, we can perform back-up in regular intervals to reduce the amount of back-up needed at a time, thereby meeting this requirement.

\subsubsection{Large state-space} PUF should be able to provide a sufficient number of unique keys throughout the lifetime of the device under protection. A strong PUF with sufficiently number of CRPs would meet this requirement by definition.

\subsubsection{Relaxed reliability requirement} The response of a PUF doesn't need to be stable over a long period of time. It only needs to be stable in successive clock cycles for one encryption and decryption. If the response i.e. the key varies in a different encryption-decryption operation e.g due to aging, that actually helps the system to be more secure by making the key unique even for the same PUF challenge, thus increasing the available keys over the lifetime of a chip.

\subsubsection{Temporary data security} Data is only valid for short period of time. The importance of the data is only valid until just after the system wakes up from sleep or a power failure. Thus we only need to provide security during this interval.


These assumptions should not be difficult to meet for the types of system for which we are trying to provide security. 
Instead of a large PUF, several smaller PUFs are activated one by one to generate a large key while not increasing the power. A reliability enhancement block is used to get clean bits for cryptographic key from the PUF response bits. The challenge for the PUF is generated from a TRNG. 

\begin{figure}
    \centering
    \hspace{-.55em}
    \includegraphics[width=3.3in]{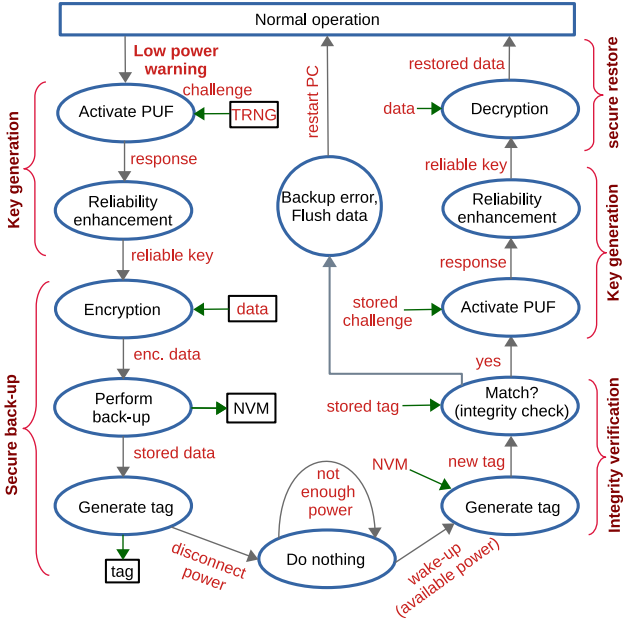}
    \caption{Flow graph for our proposed secure back-up and restore protocol}
    \label{fig:protocol_flow_chart}
\end{figure}

\section{Proposed Security Protocol}
\label{sec:protocol}

Figure \ref{fig:protocol_flow_chart} shows the overall flow of our proposed security protocol. It clearly shows the sequence of operations that a system would perform with our proposed security solution when a power failure occurs or low power warning is asserted. The minimum level of this low power warning is set by the amount of power or energy required to back-up necessary data successfully and is expected to be handled at the processor level. Figure \ref{fig:protocol_flow_chart} also shows that the key generation is required both during back-up and restore as the key is not stored anywhere and is generated during run-time. Thus, one major benefit is that we don't need to physically store the key and we can generate it whenever is needed. Moreover, for the same seed (i.e. PUF challenge), a PUF circuit would produce different keys in different ICs with the same circuit. Therefore, even if an adversary can gain access to one device, he/she would need to give the same effort to gain access to another such device. Integrity verification step is crucial as it ensures data has not been altered while the system sits on zero power. The key generation, secure back-up, and restore protocols are described below.

\subsection{Key generation}
Algorithm \ref{algo:key_gen} describes the key generation process from a PUF along with the reliability enhancement technique. As mentioned before, this involves applying a challenge to the PUF, storing responses and finally generating a clean key by applying run-time reliability enhancement technique. 


\begin{algorithm}[ht]
 \caption{Proposed secure and error-free key generation from PUF}
 \label{algo:key_gen}
 \KwResult{Cryptographic key: $\textbf{KEY}$; status: \textit{VALID\_KEY};}
 \textbf{Initialization:} \\
 \textbf{T} $\gets$ no. of samples for reliability enhancement \;
 \textbf{m} $\gets$ no. of parallel small PUF blocks \;
 \textbf{n} $\gets$ no. of response bits from each small PUF block \;
 \textbf{resp} $\gets$ \textbf{m}$\times$\textbf{n} \tcp*{total response}
 
 \SetKwFunction{FMain}{Key\_Generator}
 \SetKwProg{Fn}{Function}{:}{}
 \Fn{\FMain{}}{
  Enable(PUFs) \;
  \textbf{C} = TRNG() \tcp*{random challenge}\
  \For{$i\gets0$ \KwTo $T$}{
   \For{$j\gets0$ \KwTo $m$}{
    \textbf{$R_{i,j}$} = apply\_challenge\_to\_PUF(\textbf{C}) \;
    save\_in\_SRAM(\textbf{$R_{i,j}$}) \;
   }
  }
  \For{$k\gets0$ \KwTo $resp$}{
   \eIf{bit\_error}{
     discard\_bit(\textbf{R}[k]) \;
    }{
     \textbf{KEY}.append(\textbf{R}[k]) \;
    }
    \If{length(KEY)==nKey}{
      \textit{VALID\_KEY $\gets$ \textit{TRUE}} \;
      break \;
    }
  }
  \If{length(KEY)$<$nKey}{
    \textit{VALID\_KEY $\gets$ \textit{FALSE}} \;
  }
  \textbf{return $\textbf{KEY}$ }
 }\textbf{End\_Function} \;
\end{algorithm}

Responses are generated multiple times (twice at least) from the XbarPUF to detect the more error-prone responses bits. The main source of producing unreliable bits for an XbarPUF would be the case when two memristor's cycle-to-cycle variation are larger than their process variation. During device testing, severely unreliable bits i.e. bits with very high flipping probability, can be discarded. By applying the same challenge multiple times, we should be able to find other cases and reduce bit-error during run-time. Using temporary storage like SRAMs are used to implement this. Different challenges cause the response to depend on different sets of memristors of the crossbar and thus for different challenges, different bits of the response would cause bit-flips but they should be the same for the same challenge. In an extreme case, where the bit-error rate is so high that the no. of necessary keys bits can not be produced from the response, \textit{VALID\_KEY} token is set to false which indicates an unsuccessful key generation. 

\subsection{Secure back-up}
The data is encrypted (XORed in simplest case) with the generated key and saved in an NVM. Then a tag is generated for memory integrity purposes. This is a hardware security (HS) module that will be activated only when there is a power failure or the processor wants to go into an ultra low power mode. The sequence of operations when such a situation occurs are described in algorithm \ref{algo:secure_backup} below:

\begin{algorithm}[ht]
 \caption{Proposed secure data back-up operation}
 \label{algo:secure_backup}
 \KwData{States and other data to be backed-up, \textbf{DAT}}
 \KwResult{Encrypted data, \textbf{enc\_dat}, and \textbf{tag} in NVM}
 \textbf{Secure\_backup:} \\
 \eIf{Low-power-warning is asseted}{
  Enable(HS\_block) \;
  \textbf{KEY} = Key\_Generator(\textbf{C}) \;
  \textbf{enc\_dat} = Encrypt(\textbf{KEY}, \textbf{DAT}) \tcp*{XOR($KEY$,$DAT$)}
  write\_in\_NVM(\textbf{enc\_dat}) \;
  \textbf{tag} = gen\_tag(\textbf{enc\_dat}) \;
  save(\textbf{C}, \textbf{tag}) \;
  Disable(HW\_block) \;
 }{$Continue\_regular\_operation$ \;}
\end{algorithm}

Since memristors are non-volatile and have long retention time \cite{TaOx_Yang:2010}, the power failure phase can be long and still won't affect the data. 

\subsection{Secure restore}
When the power is back on, it needs to ensure first that there is sufficient power for the decryption of the data. Then it generates a new tag from the data and check if it matches with the stored tag for the same data. If they matches, key generation process is activated again by activating the PUF along with the reliability enhancement technique to generate a secure key. If there is a tag mismatch, the process would simply discard the data. The sequence of operation when power is back on are described in algorithm \ref{algo:restore_data}.

 

\begin{algorithm}[ht]
 \caption{Proposed secure data recovery operation}
 \label{algo:restore_data}
 \KwData{Encrypted data: \textbf{enc\_dat}, and \textbf{tag}}
 \KwResult{Restored data, \textbf{DAT} back in registers}
 \textbf{Restore\_data:} \\
 \While{P$_{available} <$ P$_{threshold}$}{
  wait \;
 }
 
 Enable(HS\_block) \;
 \textbf{enc\_dat} = read\_from\_NVM(stored\_data) \;
 \textbf{C} = read\_from\_NVM(stored\_challenge) \;
 old\_tag = read\_from\_NVM(tag) \;
 new\_tag = gen\_tag(\textbf{enc\_dat}) \;
 \eIf{new\_tag = old\_tag $\And$ \textit{VALID\_KEY} = \textit{TRUE}}{
  \textbf{KEY} = Key\_Generator(\textbf{C}) \;
  \textbf{dec\_dat} = Decrypt(\textbf{KEY}, \textbf{enc\_dat}) \tcp*{XOR($KEY$,enc\_dat)}
  write\_back\_in\_registers(\textbf{dec\_dat}) \;
 }
 {
  flush\_data() \;
  restart\_processor() \;
 }
 Disable(HS\_block) \;
 
\end{algorithm}



\section{Reliability Enhancement Technique}
\label{sec:reli_enhance}

To improve the reliability of our PUF response, we have considered two different voting approach to identify bit-flips of a PUF response and eliminate those from the final key. These two techniques, namely majority voting and all-agree voting are described below. 


\subsection{All-agree voting}
\label{subsec:all-agree-vote}


For this technique, we would apply the same challenge some `N' number of times to produce the same response `N' times. Now, some unstable bits in the response would flip and according to `all-agree' voting technique, we would only accept a response bit as part of the final key if that bit has not flipped even once in `N' evaluations. One of the benefits of this technique that this can be implemented very fairly with minimum hardware requirement, e.g. using XORs. However, even after using this technique, some bits still may remain undetected. Our target is make that probability very small.



According to this technique, a bit-error would result in a key error if a bit happens to be stable during encryption, but flips during decryption, resulting in a different key. Suppose, the probability of a particular bit being stable (stable `1' or stable `0') is p. Thus the probability of that bit being the opposite value is 1-p. Thus for N evaluation a bit being stable is $p^N$ while at least the bit flips at least once is (1-$p^N$). Now to propagate such an error due to bit-flip without being detected by this `all-agree' voting is given by equation \ref{eqn:bit-flip-prob1}.

\begin{equation}
    \label{eqn:bit-flip-prob1}
    PE_1 = p^N * (1-p^N)
\end{equation}

There is another way of error being propagated is when that bit remains stable in its less-stable state during encryption and flips during decryption. The probability of that is given by equation \ref{eqn:bit-flip-prob2}.

\begin{equation}
    \label{eqn:bit-flip-prob2}
    PE_2 = (1-p)^N * (1-(1-p)^N)
\end{equation}

Now in both of these cases, we consider the bit remains stable during encryption and flips during decryption. The same equations hold for the case where a bit flips during encryption but remains stable during decryption. Thus the probability of an error being propagated into the cryptographic key is given by the following equation \ref{eqn:bit-flip-prob_both}:

\begin{equation}
    \label{eqn:bit-flip-prob_both}
    PE_{1,2} = 2*[(p^N * (1-p^N)) + ((1-p)^N * (1-(1-p)^N))]
\end{equation}

However, this equation counts two unique cases twice. The situation is when it produces all `1's during encryption, but all `0's during decryption and vice-versa. The probability of such a case is:

\begin{equation}
    \label{eqn:bit-flip_opposite_stable}
    PE(all zeros + all ones) = 2*[(p^N * (1-p)^N))]
\end{equation}

Both equations \ref{eqn:bit-flip-prob1} and \ref{eqn:bit-flip-prob2} consider this situation and thus this is being calculated twice while calculating the overall error propagation probability as given by equation \ref{eqn:bit-flip-prob_both}. Thus by subtracting this from equation \ref{eqn:bit-flip-prob_both}, we get the final expression for a bit-flip being undetected in all-agree voting system is:

\begin{equation}
   \label{eqn:bit-flip-prob_final}
   \begin{split}
    PE (\textit{all-agree voting}) = 2*[(p^N * (1-p^N)) + ((1-p)^N * \\ (1-(1-p)^N)) - (p^N * (1-p)^N))]
    \end{split}
\end{equation}

\begin{figure}
\centering
\hbox{\hspace{-0.35in}
\includegraphics[width=4in]{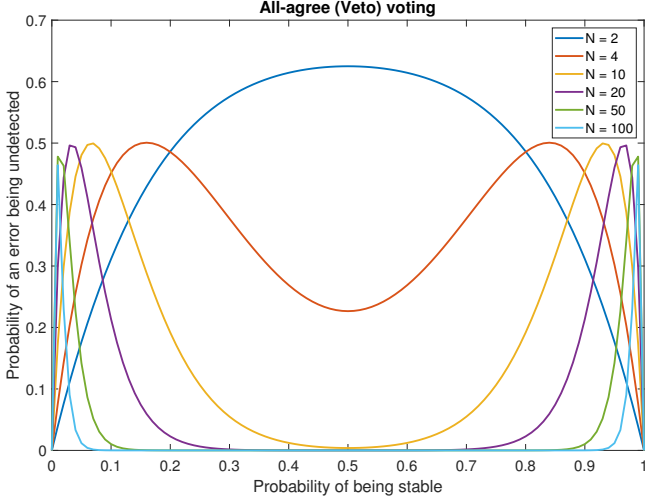}}
\caption{`All-agree' reliability enhancement technique for different number of evaluations for different bit-flip probability. It is more efficient at detecting highly unstable bits with increasing number of evaluations.}
\label{fig:all_agree_voting}
\end{figure}

We have plotted this equation in Fig. \ref{fig:all_agree_voting} for a few different `N' (no. of samples) and all different values of `p' (probability of a particular bit being stable at a binary value). 

In a PUF, there could be a few bits with a high bit-flip probability while most of the other bits should remain stable. In our particular XbarPUF, as we have shown later that out of the 32-bit PUF response, on average there are less than 2 bits with a high flipping probability while the other bits remain stable over a long period of time. Thus if we can eliminate these few bits with very high bit-flip probability, then the system would have a small probability of generating different keys during encryption and decryption and thus the need to restart the system would be less.

\begin{figure}
\centering
\hbox{\hspace{-0.35in}
\includegraphics[width=4in]{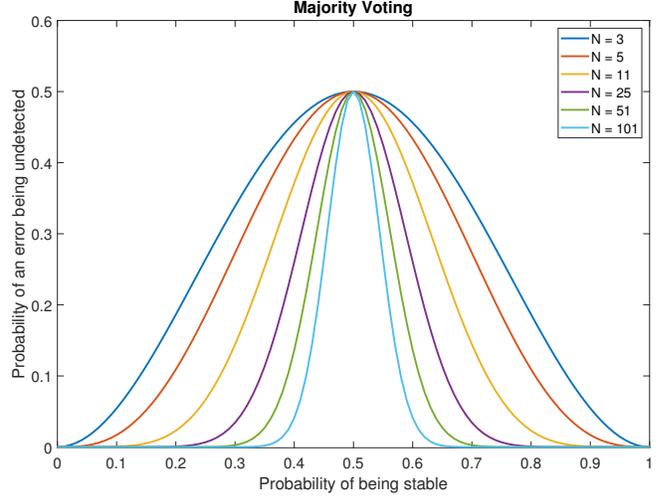}}
\caption{Effectiveness of majority voting for reliability enhancement technique for different number of evaluations and with different bit-flip probability}
\label{fig:majority_voting}
\end{figure}

From Fig. \ref{fig:all_agree_voting}, it is clear that with more number of samples (N), all-agree voting scheme is more efficient at eliminating bits with high flipping probability. On the other hand, majority voting scheme performs poorly at identifying bits with high flipping probability (near 50\%) even when a large number of samples are used, as shown in Fig. \ref{fig:majority_voting}. Both of these results are intuitive. Therefore, a good way to reduce bit-error would be to identify highly unstable bits during chip testing and keep a record for each different challenge bit. Thus, during regular operation, bits in the PUF response would have a very low probability of flipping and can be detected using a low cost all-agree voting using a very small number of samples (i.e. 2). Increasing the number of samples would increase the overhead but would be more efficient at eliminating higher bit-error rates. 

\subsection{Majority voting}
\label{subsec:majority_vote}
Majority voting scheme is a very well-known error correction technique. To employ this technique here, the same PUF challenge would be applied for some fixed number of times and a bit would be considered either a `1' if it produces `1' more than `0' among those evaluations or vice-versa. Thus unlike all-agree voting, bits are not discarded here, rather voting is used to determine their more stable binary value. The probability of determining correct binary value of a bit using majority voting technique with `N' number of samples are given by this following equation: 

\begin{equation}
   \label{eqn:bit-flip_prob_majority_vote}
   \begin{split}
    PE(\textit{majority voting}) = \sum_{r=\frac{N}{2}+1}^{N} \binom{N}{r} p^{r}(1-p)^{N-r} 
    \\= 1 - \sum_{r=0}^{\frac{N}{2}} \binom{N}{r} (1-p)^{r} p^{N-r}
    \end{split}
\end{equation}

Fig. \ref{fig:majority_voting} shows the plot of this equation for different number of samples and for different bit-flip probability. With larger number of N, the curve gets narrower i.e. it can detect more bit-flips effectively. This technique, however, is bad for bits which have high flipping probability (0.4-0.6) as the peak of the curve remains the same for any number of samples.

The overhead associated with majority voting could be high for larger number of evaluations. For example, for an N-sample majority voting, we would need counters that can count up to N for each response bit, and $log_2N$ bits of memory to store the results for each response bit as well. 


\subsection{Proposed reliability enhancement technique}

We have seen from Fig. \ref{fig:all_agree_voting} and \ref{fig:majority_voting} that all-agree voting is effective at eliminating high bit-flips while majority voting is good at detecting smaller bit-flips. Overhead increases with increasing number of samples for majority voting where all-agree voting can be implemented with using XORs and one additional bit per response bit for any number of evaluations. To reap the benefit of both techniques, we are proposing to use all-agree voting with larger `N' during chip functionality testing to eliminate bits with high flipping probability and use majority voting with small `N' during run-time to reduce bit-error. If our all-agree voting can eliminate highly unreliable bits beforehand, then majority can be used with a very few samples during run-time, thereby also ensuring lightweight operation. However, in case where there are some bits with high flipping probability, all-agree voting should be used as shown in our hardware implementation later.

Depending on the bit-flipping profile of a particular PUF, the overhead associated with the implementation of either all-agree or majority voting schemes along with the allowable power, area, delay constraints and security requirements of a system, we can choose a particular reliability enhancement technique over the other. The idea is not to eliminate bit-error altogether, but to reduce its probability to such extent that the cost or overhead associated with full system restarts (when there is a key error) would be overcome by the smaller overhead associated with that particular technique.

\begin{figure}
\centering
\hspace{0em}
\includegraphics[width=3.75in]{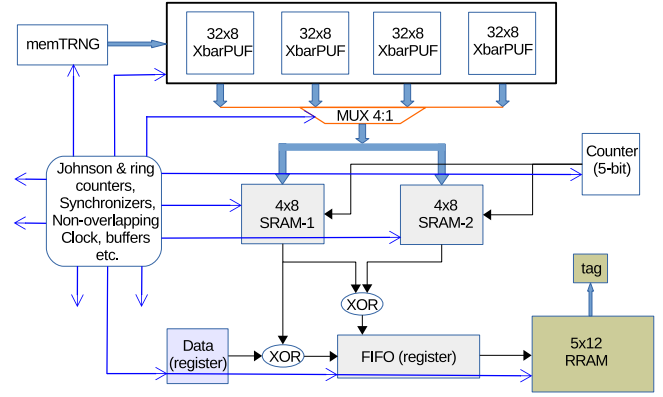}
\caption{System block diagram showing different components of our proposed hardware security design.}
\label{fig:system-circuit-diagram}

\end{figure}

\section{System Implementation}
\label{sec:system-design}

One of the main contributions of this work is that we have implemented our proposed security protocol at circuit level to get more realistic idea about the implementation complexity and system overhead. Specifically we have designed our system at transistor level with 65$nm$ CMOS technology in Cadence Virtuoso. Thus all our data, including security metrics and resource overhead are calculated at this level, thereby giving us more accurate estimation of real-world implementation. This prototype system generates a 16-bit key from a 32$\times$32 XbarPUF. Fig. \ref{fig:system-circuit-diagram} displays the different components of our proposed system and these are described in the following subsections.


\subsection{True random number generator (TRNG)}
\label{subsec:trng}
We need to provide random, favorably unique challenge to the PUF during each back-up to generate a unique response. Thus a random challenge would act as a random seed for this key generation block. For that purpose, we are proposing to use a true random number generator (TRNG) to produce random challenge vectors. TRNG output should not be affected by environmental changes as that might compromise the randomness of the TRNG and thus the security of the whole system. Deploy-able embedded device might go through various environmental changes. Thus the TRNG needs to be robust against environmental changes. Researchers have shown good TRNG built from memristor \cite{Jiang:2017_Memristor_TRNG}. In this work, we propose to use the memTRNG introduced in \cite{Mesbah_TRNG:2019}. The required area of this TRNG is small, compatible for our design as it also uses memristors to generate random numbers, and importantly, robust against temperature and supply voltage change as analyzed in \cite{Mesbah_TRNG:2019}.

\subsection{Time-multiplexed XbarPUF}
For a memristive crossbar PUF, the major contributing factor to total power consumption is its RESET phase \cite{Mesbah:2016} when all the memristors are RESET from LRS to HRS and static current flows through the memristor crossbar. Response generation is parallel in the XbarPUF and, therefore, with increasing number of response bits, there is a linear increase in static current as well. Thus power consumption in RESET phase might be the bottleneck of the system by having a larger peak power demand. To reduce this power, we have divided up the XbarPUF into separate smaller PUFs, each working on different times, generating one portion of the response vector. For this prototype design, the 32$\times$32 XbarPUF is divided up into four 32$\times$8 XbarPUFs. Depending on the system requirement and resource available, this number would change. A state machine with four states for these four PUFs is designed so that on each state, a single PUF is being RESET, challenge is applied and response is generated and saved in a temporary memory. This also implies that it would require 4$\times$ more clock cycle to generate all the responses compared to a single 32$\times$32 PUF, but the average power during RESET is also reduced approximately by a factor of four. Area is slightly increased due to the need of separate read-write-circuitry for each XbarPUF block. Each XbarPUF design is identical and implemented as shown in Fig. \ref{fig:xbarpuf} \cite{Mesbah:2016,Mesbah_SA:2018}.

\subsection{Reliability Enhancement Block}
We have introduced the theory and rationale behind implementing our reliability enhancement block. For our demonstration, we are using a 32$\times$32 PUF to generate a 32-bit response from which a 16-bit key is extracted. The amount of extra bits to be used can be estimated by simulating the reliability of the PUF in its worst case operating condition. We are implementing a simple two-count majority/all-agree voting system here using two SRAMs. A single unique challenge is applied to the PUF two times and the two sets of response vectors are generated and saved in two different temporary memory i.e. SRAMs. One SRAM holds the PUF response generated in the first clock cycle while the second one holds the response generated in the second cycle. We have designed the two SRAM blocks ourselves, each sized 4$\times$8 to hold 8-bit response from each of the 4 PUFs. SRAM cells are sized appropriately to ensure minimum delay and area as well as minimum chance of a read or write error. For a complete system, existing volatile memory can be used for this purpose, thus requiring no extra hardware. To increase the confidence over the reliability of the key, we can repeat these steps over multiple cycles for the same challenge to have more information about bit-flip, at the expense of larger overhead.

\subsection{Stable Key Generation and Cryptographic Block}
During key generation phase, contents from the two SRAMs are compared bit by bit and the first 16 reliable bits are taken as the key. A counter is used to keep track of the number of reliable bits and once that count reaches 16, it stops and the 16 `clean' bits are ready to be used as key. To reduce delay, a whole row of 8-bit data is written at a time. Thus it takes 4 clock cycles to write all 32 response bits from the 4 XbarPUFs. The second SRAM takes another 4 cycles. Since the data access to RRAM is 1 bit at a time, the two SRAM are also read bit by bit. Two bits from the same location of the SRAMs are read at a time, XORed to check if they are same or not and only used as part of the key if they are the same. Then one bit from the data is XORed with this valid key bit and can be temporarily saved in a FIFO (first in first out). Finally data from the FIFO can be written into the non-volatile memory which is RRAM in this case. The 16-bit XOR block is used as the both encryption and decryption block. Thus our proposed system implements a homomorphic encryption. If the system can allow more resource overhead, then we can swap this XOR based encryption with any robust encryption system using that same key. Moreover, instead of comparing bit by bit, we can also compare multiple bits at a time to increase the speed of back-up operation, although speed gain is limited by the memory access speed.

\subsection{RRAM as Non-Volatile Memory}
We have used memristors to design completely two different circuits in this work. First one is to build the XbarPUF. But we are also using memristors as the NVM of our system. We need a minimum of 16-bit RRAM to hold 16-bit encrypted data. However, to keep things generic, we are assuming that this NVM is being used for holding other information as well and thus we are using a larger 5$\times$12 RRAM. The HRS and LRS of a memristor are considered as logic `0' and logic `1', respectively. This RRAM is an 1T1R structure i.e. each memory cell consists of a memristor and an access transistor. The sizing of this access transistor depends on the HRS and LRS values of a memristor to maximize read-write noise margin but minimize power/area/delay. More details about this RRAM design are skipped for brevity.

\subsection{Sneak-path based Tag Generation for Data Integrity Verification}
\label{subsec:memory_integrity_check}
The purpose of the PUF based encryption is to provide confidentiality to the system. It prevents an attacker from reading out sensitive information from the system. However, it doesn't provide data integrity and therefore any modification in the non-volatile memory in the sleep mode would be unnoticed. A lightweight integrity checking system targeted for memristive memory is used here in order to detect unauthorized modification in the system information saved in the memory. 
This integrity checking scheme leverages the sneak path current based tag generation from a crossbar memristive memory. Sneak path currents in a crossbar memristive array are read using multiple columns and converted to digital bits in this tag generation method. Details about this system can be found in \cite{Majumder_IntegrityCheck:2018}. This scheme does not require an additional hash function or message authentication code (MAC) since  the memory itself is used as the tag generator. It has been therefore proposed as a security primitives for resource constrained systems \cite{Mesbah_CEM:2018}. This data or memory integrity scheme would also help to detect any data error or unreliable key bits. This data integrity check is very crucial for our system as it would be able to detect any data error, key error or unauthorized modification of backed-up data during low power mode of a processor. During wake-up, it calculates a new tag and compare with the stored tag. If these two tags do not match, then the processor would reject the back-up data and restart the whole computation.


\subsection{State Machine and Control Logic Design}
\label{subsec:state}
We have used Johnson counters and ring counters to control different areas of the circuit and to design the whole state diagram for our proposed system. For the case of a resource constrained system, this security block is only enabled whenever there is a low power warning or whenever system is recovering from a power failure. For regular embedded processors, this security block would be enabled just before a sleep and just after a wake-up to allow for encryption or decryption, respectively. In first two states, the XbarPUF is activated twice to produce two sets of 32-bit responses using the same challenge. To reduce the peak power consumption during a whole matrix of memristor write in these states, the XbarPUF is divided into four smaller PUFs as mentioned before. Therefore, in each first and second state of the system, there are four sub-states, one for each smaller XbarPUF block. Responses from each smaller XbarPUF are saved in two SRAMs. Each of these sub-states again has three different phases: Reset-all, challenge, and Read to generate a response vector from an XbarPUF block. The next few states together are used to read from these SRAMs one bit at a time and then compared with each other using an XOR block. If they match, the corresponding bit is saved in the RRAM. In the next state, the tag generation control block is activated and a tag is generated from the RRAM. This tag needs to be saved and is stored in a secure memory. 

During wake up, a new tag is regenerated from the RRAM data and compared with the stored tag. If they do not match, the system sends an alert to the processor, that the system might be compromised and all backup data are discarded. But if the tags match, then the PUF is activated again and the response and key are regenerated. Then data from the RRAM is read bit by bit, XORed with the corrected PUF responses (enc-dec key) from the two SRAMs to get the original decrypted data bit. This process can easily made faster by allowing multiple bit (i.e. a byte at a time) read-write access into the RRAM.

It is very important to ensure that no two states are active at the same time even for a very short period of time because that might enable multiple security blocks or PUFs. Non-overlapping clock generation circuits are used at the output of state decoders to prevent this situation arising from slow transitioning signals which might create metastability in the design. Two flip-flop based synchronizers are also used to ensure realiable data transfer from one clock domain to another where both clocks are of the same frequency and constant phase difference in this application.

\section{Probable Attack Scenario}
\label{sec:probable_attacks}
\subsection{Malicious read}
This is the main motivation behind this work. An attacker might try to read the contents from the backup non-volatile memory and thus gain sensitive information. Our PUF-key based one time pad (OTP) encryption scheme encrypts the data to prevent direct interpretation of sensitive information. As we have explained before, we are effectively implementing a OTP here. OTP is theoretically the most secure encryption if we can fulfill its requirements: (1) random key, (2) key changed on during each encryption, and (3) key as large as the data. Our key is random as it comes from a PUF and in a small embedded system where amount of backup is small, key can be made as large as the data. Moreover, since the backup operation is infrequent and the state-space of a strong PUF is large, a PUF key is unlikely to be repeated in a practical time-frame and thus replay attack is improbable. Key sharing is another weakness an OTP implementation which is not a concern here as we are not communicating with outside world using this key. Therefore, we are fulfilling all the requirements of an OTP and ensuring maximum security with a unique random key.

\subsection{Malicious write}
Instead of trying to read the information, an attacker might try to change the contents of the RRAM arbitrarily, thereby creating an erroneous calculation. Data error or key error can also result in an error, especially during power failure and time sensitive back-up operation. Sneak-path based one way memory integrity checking should be able to detect any kind of data alteration. During wake-up, the program calculates the tag from the backup memory and check with its saved tag. If they do not match, the processor simply rejects the backup data and restarts the operation. This would increase the overhead as processor can't make use of previous states and data, but it makes the system robust against any erroneous or harmful procession of information.

\subsection{Readout or alteration of the PUF challenge or secure tag}
The challenge to generate the PUF response and the tag are saved in a secure NVM, assumed inside the CPU of the embedded processor. Here, we assume that attacker wouldn't have access to these small bits of secure memory. If the PUF is robust against modeling attack, as shown in \cite{Mesbah_TNANO:2017} for XbarPUF, just getting access to the challenge shouldn't compromise the security. Moreover, continuous access to the PUF is not permitted in this system as the PUF is only used at some certain stages. Finally, illegally changing the PUF challenge would almost definitely change the PUF response and, therefore, the encryption key which would result in an erroneous data. Pre-computing a tag or hash with the encrypted data would help to prevent such scenarios and detect alteration in either data or key.

The produced tag is calculated from the sneak path currents of the crossbar of memristors \cite{Majumder_IntegrityCheck:2018}, i.e. this is an analog in-memory computation. Because of the analog nature of memristors and its die-to-die and cycle-to-cycle variation, it should be very difficult to repeat a memristor's exact resistive state to regenerate the same sneak-path current and the same tag. Thus once the data memory and tag are written, any attempt of writing should corrupt the data and the tag and thus processor would be able to detect it easily.

\begin{figure}
\hbox{\hspace{0.25in}
\includegraphics[width=3in]{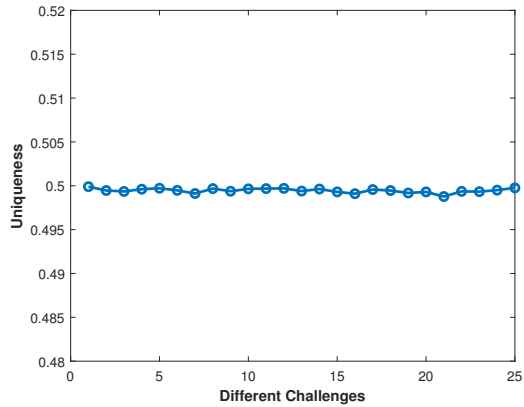}}
\caption{Average uniqueness results for 500 different chips. Even for different challenges, this value is very close to ideal value of 0.5}
\label{fig:uniq}
\end{figure}

\subsection{Modeling attacks}

One of the concerns of PUFs, especially those which can be modeled as an additive linear delay model (like arbiter PUF), are susceptible to machine learning based modeling attacks \cite{Ruhrmair:2013}. The idea is that an attack would gather a subset of challenge-response pairs from a strong PUF and use that small subset, would build a machine learning model that can predict responses for unknown challenges. This effectively reduces the possible number of useful keys from a strong PUF. In previous works, it was shown that memristor based PUFs are also vulnerable to machine learning modeling attacks. However, unlike authentication applications where a PUF is requested to generate a few different responses from different challenges to compare with the stored database for that PUF, the PUF in our proposed system would not give any access to its keys during its operation because it won't be saved anywhere. 

In an extreme scenario, an adversary can gain control of a device, cause it to do back-up and restore continuously with known data value so that it can gain information about the responses by doing a malicious read of the back-up data and challenge bits. To prevent this from happening, a PUF should be able to resist modeling attacks. A simple and lightweight modification is to introduce response bit XORing and column shuffling technique which can drastically reduce the modeling accuracy of an XbarPUF, thereby increasing the robustness against machine learning based modeling attacks. These techniques are discussed in details in \cite{Mesbah_TNANO:2017}. Moreover, any device with our proposed system should also have a tamper detection mechanism so that the expected number of back-up in a given time doesn't exceed a certain value. This would prevent an attacker from building a database in a short period of time.

\section{Security Properties}
\label{sec:security_properties}

As we have mentioned before, we have implemented a 32$\times$32 XbarPUF as the key generator. First, we have evaluated this PUF in terms of several security metrics listed below \cite{Maiti:2013}.

\begin{multicols}{3}
\begin{itemize}
    \item Uniqueness
    \item Uniformity
    \item Bit-Aliasing
    \item Diffuseness
    \item Reliability
    \item Steadiness
\end{itemize}
\end{multicols}

\begin{figure}
\hbox{\hspace{0.25in}
\includegraphics[width=3in]{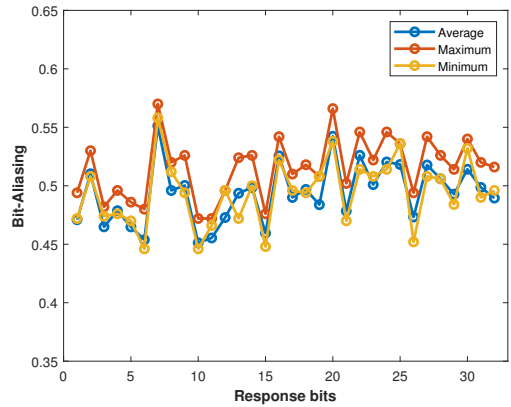}}
\caption{Summary of results for bit-aliasing for all 32 bits from 500 different chips. The minimum and maximum value for each bit are also shown.}
\label{fig:bitali}
\end{figure}

These metrics are defined and explained elaborately in \cite{Maiti:2013}. Interested readers are encouraged to learn the formal definitions and equations for these metrics from there. Uniqueness measures how different the responses are from one chip to another for the same challenge. Bit-aliasing determines the evenness of 1's and 0's in a each bit position of a response for different chips. Thus both of these metrics evaluate a PUF's response across different devices. To evaluate uniqueness and bit-aliasing, we have performed Monte Carlo analysis for 500 different chips, each with 25 different challenges. Uniqueness results are shown in Fig. \ref{fig:uniq}. Our designed XbarPUF displays near ideal (50\%) uniqueness value for all these different challenges which shows the strength of this implemented system as an unclonable hardware security module. The results for bit-aliasing for each of the 32 bits of a response vector for different challenges are shown in a box-plot in Fig. \ref{fig:bitali}. For all bits, the bit-aliasing value is within the [0.45 0.55] window and the average bit-aliasing of all these bits is 0.5, very close to its ideal.

\begin{figure}
\hbox{\hspace{0.1in}
\includegraphics[width=3.2in]{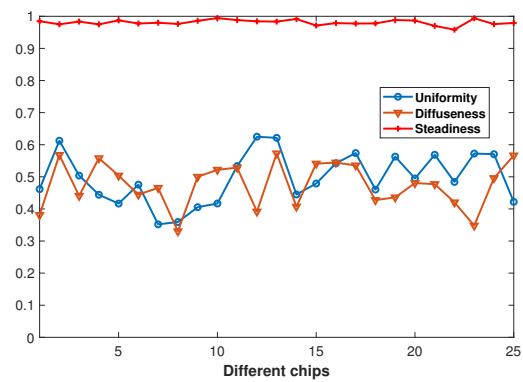}}
\caption{Summary of results for average uniformity, diffuseness and steadiness from 500 different challenges and for 25 different chips.}
\label{fig:unif_diffuse_steady}
\end{figure}


Uniformity and diffuseness measure a PUF's performance across the challenge space. For the same chip, if different challenges are applied, the responses should be as different as possible and ideally these metrics should be equal to 0.5. Steadiness is a metric that evaluates the stability of a PUF's response. It represents bias of individual response bits of a PUF on average. Specifically, it measures the degree of bias of a particular bit towards a `1' or `0' over many different cycles as defined in \cite{Maiti:2013}. To evaluate uniformity, and diffuseness, we have run Monte Carlo simulation for 500 random challenges, for 25 different chips. We have evaluated steadiness for 25 different chips for 500 cycles. Results for uniformity, diffuseness, and steadiness are shown in Fig. \ref{fig:unif_diffuse_steady}. As it can be seen, although for different chips, these numbers deviate from this ideal value, they do not show very large deviation and display an average uniformity of near 0.5 and an average diffuseness of near 0.5. This shows the applicability of this design as a strong PUF, capable of generating unique keys for different challenges. Also from this figure, it can be seen that steadiness for these chips are very close to the ideal value of 1.

\begin{figure}
\hbox{\hspace{-0.05in}
\includegraphics[width=3.3in]{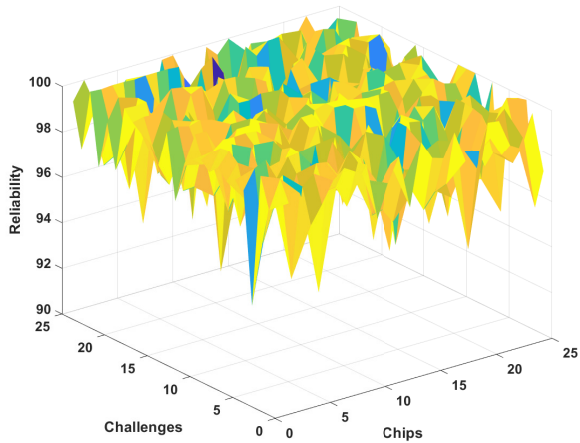}}
\caption{Detailed reliability results generated from 500 different cycles for 25 different chips and 25 different challenges.}
\label{fig:reli}
\end{figure}

One of the major concerns around PUF is their reliability. Because of the way how a PUF operates to generate device-specific signature from tiniest process variation, any small change in environment or the presence of noise can make a PUF's response prone to change undesirably. Reliability, closely related to stability (distinction is explained in \cite{Maiti:2013}) are used to describe how reliable a PUF's response is when a same challenge is applied again and again \cite{Maiti:2013}. Because, reliability might be the most important and most concerning metric to evaluate a PUF, we have shown a detailed reliability result using Monte Carlo analysis for 500 different clock cycles, for 25 different challenges in 25 different chips. This result is shown in Fig. \ref{fig:reli}. This 3-D plot shows that for all chips and all different challenges, the XbarPUF shows at least 92\% and on average 98\% reliability. However, it is to be noted that this result is for the XbarPUF directly, before applying any reliability enhancement technique. Using our proposed reliability enhancement technique, the idea is to take only reliable bits from the response. 

After evaluating the security of the PUF, now we are interested in evaluating the security of the tag generation method that we have used in this work. This is a sneak-path based tag generation, specific to RRAM or memristive memory and the details can be found in \cite{Majumder_IntegrityCheck:2018}. Three metrics to evaluate any tag generation or hashing method are: uniformity (different than PUF's uniformity), diffusion, and avalanche effect. Uniformity dictates that the probability of each tag bit being either 0 or 1 should be equal to each other. To fulfill the diffusion property, if 1-bit of the data is changed, each tag bit should have a probability of being flipped to be equal to 0.5. Avalanche effect means the property of the output tag to be very different for two very similar data, maybe differing by just 1 bit. This means even if 1-bit of data is changed, almost half of the tag bits should be changed. Table \ref{tab:tag_generation} presents the results for this tag generation method. This result is slightly different from the one presented in \cite{Majumder_IntegrityCheck:2018} as it is regenerated for the memristor type and crossbar used in this work.

\begin{table}
\centering
  \caption{Security properties of the tag generation method}
  \centering
  \label{tab:tag_generation}
  \begin{tabular}{cccl}
     \toprule
     Tag size & Uniformity & Avalanche & Diffusion\\
     \midrule
     6 &0.9869& 0.4953& 0.4971\\
     8 &0.9637 &0.4951 & 0.5062\\
  \bottomrule
\end{tabular}
\end{table}

\begin{figure}
\hbox{\hspace{-.1in}
\includegraphics[width=3.8in]{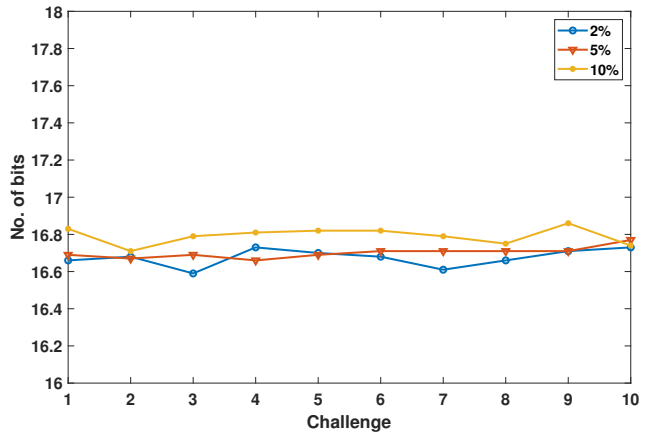}}
\caption{Figure shows the average number of bits to produce 16 `good' bits from 50 different chips. The results are generated for three different cycle-to-cycle variations and a $50^0$C temperature change.}
\label{fig:avg_clean16bits}
\end{figure}


\section{Security Evaluation}
\label{sec:security_results}
\subsection{Malicious read}
An attacker can read stored back-up data and can gain sensitive information. Since our data is encrypted, this would increase the complexity to learn anything from the data. We have implemented an OTP or one time pad here. Since an OTP key is the same length as the data itself and random, for a 16-bit data, the whole space of 2$^{16}$ possible combinations are equally likely to be a key. Thus OTP is not vulnerable against brute force attack because the attacker doesn't gain any new information from a brute force on a data encrypted using OTP, as explained in \cite{OTP_wiki:2020}. For example, with an N-bit key, the possible number of key combinations, N$_{key}$ before and after brute force attacks are:

\begin{equation}
\label{eqn:brute_force}
\begin{split}
    N_{key, before} = 2 ^ N \\
    N_{key, after} = 2 ^ N \\
\end{split}
\end{equation}

Thus for any two different data, d$_1$ and d$_2$ in all possible data space D, the probability of any ciphertext c being equal to either d$_1$ or d$_2$ is:

\begin{equation}
\label{eqn:OTP}
\begin{split}
    P[(d_1 \epsilon D] =c) = P[(d_2 \epsilon D)=c] \\
\end{split}
\end{equation}

Since brute force doesn't add any new information which wasn't already available to an attacker, OTP can maintain perfect secrecy.

\subsection{Malicious write}
\begin{figure}
\hbox{\hspace{0.3in}
\includegraphics[width=2.8in]{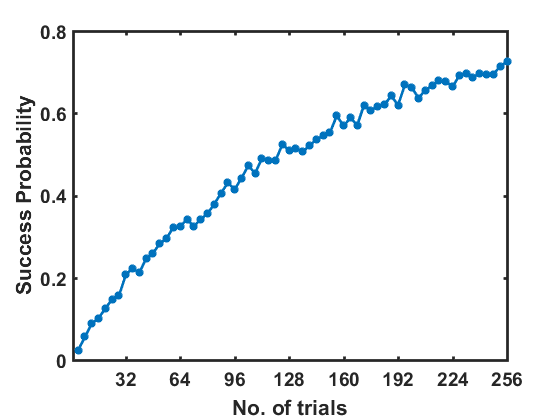}}
\caption{Probability of success in a spoofing attack with number of trials. The more trials an attacker can perform, the higher the chance that the data matches the tag. No. of effective trial is 1 in this protocol since the tag is updated on each cycle.}
\label{fig:collision}
\end{figure}

Another security goal for this work is to verify integrity of the backed up data on power regain. When the system goes to a critically low power stage, system state is saved to a non-volatile memory. An attacker may be able to launch a spoofing attack by connecting the non-volatile memory from a different source in the network in order to perform malicious write. However, the integrity verification method described earlier should be able to detect any offline modification to the memory by generating a tag. Attacker's goal in the spoofing attack is to modify the memory content in such a way that it matches the tag. The probability of being successful in this attack depends on the uniformity of tag distribution and also the number of trials attacker perform before the memory is updated by the authorized user. The tag generation method exhibits a uniform distribution for generated tags. Therefore, we analyze the probability of a successful spoofing attack as a function of number of trails as shown in Fig. \ref{fig:collision}. As it can be seen that the probability increases with the number of trials. For a given trials, the probability depends on the tag size. A hypothetical scenario that an attacker may leverage is that the back up data stored in the non-volatile memory was not changed for few cycles of power down and regain. In that case, an attacker can perform more spoofing trials on the data in order to be successful where the data matches the tag. However, the tag generation protocol randomly reconfigure its reserved bits as a timestamp before every back up stage and generate a new tag even if the data is not changed at all. Therefore, the effective number of trials for an attacker in order to perform a spoofing attack is 1. The success probability for a 8 bit tag with a single trial is nearly 1/256. This is sufficient due to the consideration that the tag is updated on each power down stage regardless of the data and the attacker cannot get multiple trials on guessing a data-tag pair. Due to the same reason, this protocol can also prevent replay attack where an attacker remembers the tag on a past data and replace the present (data,tag) pair with the past one. Each data has a large number of variants for the tag depending on its timestamp. A tag on a data at one instant would be completely different than the tag on the same data at a different instant.

\subsection{Modeling attacks}
In this work, we are assuming that the attacker can read any non-volatile data which means he can gain access to the stored challenge vector as well. Since the response of a PUF depends on the tiniest variations caused by unintentional manufacturing variation and, thus, is very difficult to predict. However, researchers have shown that with the help of machine learning techniques, PUFs can be modeled too using only a subset of the total CRPs \cite{Ruhrmair:2013,Hospodar-modeling:2012}. In a previous work, we have also shown how a 32$\times$2 XbarPUF can also be modeled and predicted with high accuracy and shown how to apply mitigation techniques \cite{Mesbah_TNANO:2017} to reduce that accuracy significantly. Here, we are recreating the work with the memristor models and circuit parameters that we have used in this work. We have used machine learning toolbox from python scikit-learn \cite{scikit-learn} and presented the result against four different models here in Table \ref{tab:modeling_attack}. We have collected 5000 CRPs from an abstract model of our XbarPUF, developed in \cite{Mesbah_TNANO:2017} and then used two-third of the data for training and the rest for testing. It is clear that the accuracy is almost like a random guess of a coin flip (50\%) for all four models, namely support vector machine (SVM) with Gaussian kernel (RBF), logistic regression (LR), naive Bayes Gaussian, and AdaBoost ensemble. The mitigation techniques, XORing and column swapping are discussed in \cite{Mesbah_TNANO:2017}.

\subsection{Readout/alteration of secure information}
PUF challenge and generated tag from backup data also need to be saved in NVM. Besides read/write of regular backed-up data, we are pessimistically assuming that an attacker can also gain access to these sensitive information. Now without a good prediction model, this challenge wouldn't reveal any information about the back-up data. As we just saw that the prediction accuracy using modern ML models against our PUF is random, this wouldn't reveal anything about the data itself. Moreover, changing the challenge would change the response vector significantly which are confirmed from the very good uniformity, bit-aliasing values of this PUF. Because of the good collision property of our tag, it won't be easy to find the data even if the tag is known. Finally, because of good avalanche property as shown in Table \ref{tab:tag_generation}, even if a single bit of tag is changed, almost half of the data bits would change, thus making it robust against such malicious changes.

\begin{table}
\label{tab:modeling_attack}
\caption{Results from modeling attacks for an XbarPUF applied for different machine learning algorithms}
\centering
\begin{tabular}{|c|c|c|c|c|c|c|c|}
    \hline
	 \multicolumn{2}{|c|}{\textbf{SVM (RBF)}} & 
	\multicolumn{2}{|c|}{\textbf{L. R.}} & 
	\multicolumn{2}{|c|}{\textbf{Gauss. N. B.}} &
	\multicolumn{2}{|c|}{\textbf{AdaB. Ensem.}}
	\\
	\hline
	Train & Test & Train & Test & Train & Test & Train & Test\\
	\hline
     51.27 & 49.48 & 56.17 & 50.16 & 56.31 & 50.25 & 56.16 & 50.24 \\
    \hline
\end{tabular}
\end{table}

\begin{figure}
    \centering
    \includegraphics[width=3.45in]{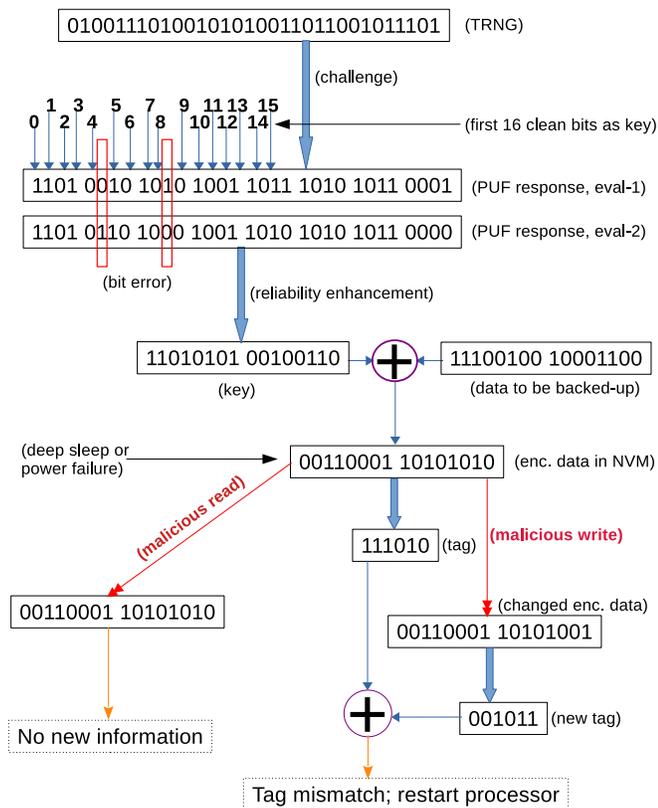}
    \caption{Demonstration of malicious read and write attacks on an NVM during power down or sleep. Attacker can read data maliciously but it won't give him any new information about the data which wasn't already available to him. Malicious write on the other hand is detected by the integrity checking method as the altered data would produce a different tag. This causes in a data rejection and processor restart which results in a data loss but malicious data doesn't enter the system.}
    \label{fig:attack_demo}
\end{figure}

Figure \ref{fig:attack_demo} shows a visual demonstration of an attacker trying to perform malicious read or write on an NVM. Maliciously reading the data won't reveal anything to an attacker, even by performing brute force attack, as the data is encrypted beforehand using our OTP encryption. Altering the data maliciously would change the tag generated from the encrypted NVM and since the old stored tag during back-up is compared with a new tag generated during restore, this integrity checking method would be able to detect any malicious write. This means the data won't be reused, processor loses information, but an attacker cannot inject any maliciously altered data into the IoT device processor.

\section{Performance Evaluation}
\label{sec:performance}

\begin{figure*}
\centering
\hbox{\hspace{-.99in}
\includegraphics[width=8.9in]{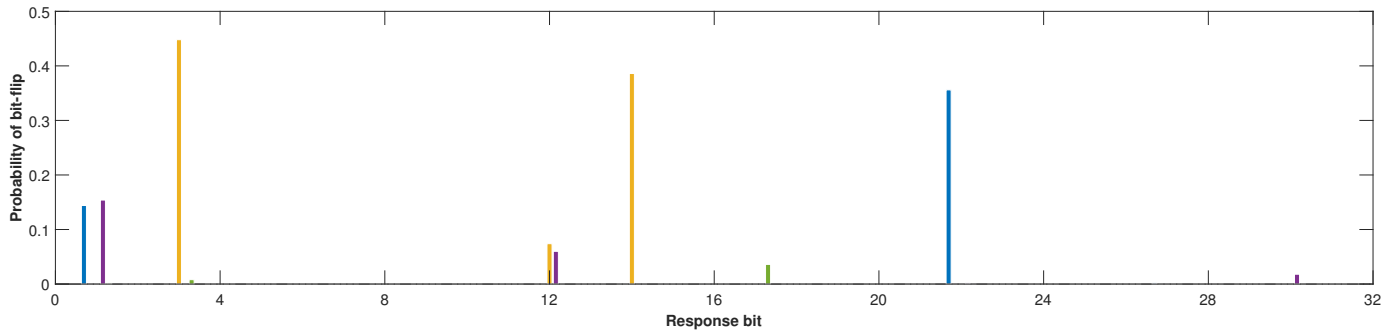}}
\caption{Probability of bit-flip for a same chip for 10 unique challenges. Plots showing the bit-flip probability for all 32 bits in a response, evaluated for 500 cycles. Different challenges cause different bits to flip i.e. there is no single set of globally unreliable bit.}
\label{fig:bit-flip-prob}
\end{figure*}

We are generating a reliable cryptographic key from the PUF response. We are accepting the fact that there would be unreliable bits in the PUF response and that we'd get at least the required number of reliable bits (16-bit here) from the PUF response (32-bit). These extra bits present an overhead to the system. To measure the minimum number of such extra bits, we have run Monte Carlo simulation to predict this situation. Memristor's cycle-to-cycle variation is the root cause of unreliability in the XbarPUF. In cases where the the cycle-to-cycle variation of a pair of memristors dominate over their process variation, then that pair would have a higher bit-flip probability in different cycles. Moreover, since our main application here is deploy-able IoT or embedded system, we also have to consider rapid environmental changes. To emulate this situation, we have considered the case when temperature changes drastically from room temperature to $50^0$C above room temperature between successive clock cycles. A change in supply voltage would change the switching speed of memristors but since we are using a large enough switching time to ensure complete state transition, this voltage change shouldn't affect the final memristive values and the PUF response. Thus we have considered only cycle-to-cycle variation and temperature change for this analysis.

Fig. \ref{fig:avg_clean16bits} shows the average number of bits it takes to produce 16 `clean' bits to be used as the cryptographic key for three different cycle-to-cycle variation of memristors. As cycle-to-cycle variation is the main culprit behind producing unreliable responses from this XbarPUF, we have used three different sets, 2\%, 5\%, and 10\%. From fig. \ref{fig:avg_clean16bits}, we can see that with increasing number of cycle-to-cycle variation, the minimum no. of required bits increases. However, because of the robustness of our XbarPUF design, the amount of bit loss is small ($\approx$0.8) even for a 10\% cycle-to-cycle variation of memristance, with 50$^0$C temperature change. The goal of our system is to produce 16 error-free bits in two consecutive cycles of encryption and decryption.

To get an idea about the amount of bit-flip, we have applied different challenges in the same chip, each for 500 times. The percentage of time that the bits flipped are shown in Fig. \ref{fig:bit-flip-prob} for 10 such challenges and for a particular chip using Monte Carlo analysis. As we can see from this figure, there are one or two bits per challenge which might have a much larger bit-flip probability and can be detected and discarded during functionality testing of the chip. The purpose of our reliability enhancement block is to mitigate the impact of bit-errors that remain undetected during testing and might cause a key-error during run-time using all-agree voting scheme. From this figure, we can also see that our XbarPUF based key generation method is able to produce 30 `clean' key bits on average from a 32-bit response. To account for higher bit-errors in some chips, the ratio of no. of key bits with no. of response bits may be reduced which would increase the yield of the design.

\begin{table*}
  \centering
  \caption{Total area/transistor count for different components of the system}
  \label{tab:are_overhead}
  \begin{tabular}{cccl}
    \toprule
    Design block & Component & Count& Comment\\
    \midrule
    XbarPUF & Memristor & 64$\times$16 & 10$nm$ $\times$ 10$nm$\\
    $\times$4 & Sense amplifier & $\times$8 & 9-T cell (base width 1.2 $\mu m$) \cite{Mesbah_SA:2018}\\
    & Row controller & $\times$64 & 8 pass-gates\\
    & Column controller & $\times$8 & 10-T 2-R\\
    \hline
    Encryption-Decryption block & XOR & $\times$ 16 &\\
    \hline
    Tag generation block & MUX2to1 (regular) & 3$\times$12 & 120$nm$ NMOS and PMOS\\
    & MUX2to1 (wide) & 3$\times$12 & 12$\mu m$ NMOS and PMOS\\
    & Sense Amplifier & $\times$12 &9-T cell (base width 1.2 $\mu m$) \cite{Mesbah_SA:2018}\\
    & Resistor & $\times$12 & load resistor = $\sqrt{HRS*LRS}$\\
    \hline
    memTRNG & Memristor, NMOS & $\times$2 & (10$nm$ $\times$ 10$nm$)\\
    & Differential op-amp & $\times$1 & 5-T cell (base width 1.2$\mu$m) \\
    \hline
    RRAM & 1T1R & 5$\times$12 & 1 memristor (10$nm$ $\times$ 10$nm$), 1-NMOS (4.8$\mu m$)\\
    & Sense amplifier & $\times$12 & 9-T cell (base width 1.2 $\mu m$) \cite{Mesbah_SA:2018}\\
    & Pass gate & $\times$12 & large (12$\mu m$) NMOS \& NMOS\\
    & Decoders & $\times$1 & 4to16 \& 2to4 decoders \\
    \hline
    SRAM & SRAM cell & $\times$32 & 6-T cell (120$nm$,240$nm$)\\
    $\times$2 &  Pre-charge & $\times$8 & 3-PMOS (1.2$\mu m$)\\
    & Sense Amplifier (SA) & $\times$8 & Current Latched SA (9-T) \cite{Kobayashi:1993}\\
    & Column buffer & $\times$8 & 4-T (1.2$\mu m$), 2-NOT\\
    & Address decoder & $\times$8 & 2to4 \& 3to8 decoder\\
    & Basic gates & $\times$8 & (AND, OR, NOT); min. width \\

    \hline
    Others & Counters, Buffers, flip-flops, &&\\
    &Non-overlapping clock generator etc. &&\\
    &state decoders, basic gates etc. &&\\

  \bottomrule
\end{tabular}
\end{table*}

\begin{table}
\centering
  \caption{Power consumption of the system in different stages of operation (State 1 and 2 involve a reset of 64$\times$64 memristors at once and thus have large static current)}
  \label{tab:power_overhead}
  \begin{tabular}{ccl}
     \toprule
     State & Average current ($\mu$A) & Comment\\
     \midrule
     State 1 & 143.9 & PUF response generation\\
     && + SRAM-1 write\\
     State 2 & 151.8 & PUF response generation\\
     && (again) + SRAM-2 write\\
     State3to10 & 0.143 & Both SRAM read \\
     &&  + RRAM write \\
     State$11$ & 3.28 & Tag generation\\
     Total & 24.80 & Overall average current \\
  \bottomrule
\end{tabular}
\end{table}

The area, power, and delay overhead for different components and in different phases of the system are shown in Table \ref{tab:are_overhead}, \ref{tab:power_overhead}, and \ref{tab:delay_overhead}, respectively. From table \ref{tab:are_overhead}, it might seem like the required area overhead of our proposed system is large. However, if we take a closer look at this table, we can see that except for XbarPUF blocks, most other units like SRAM, decoder, RRAM, counters etc. are actually parts of a regular processor and memory and can be reused for our proposed security implementation. We have designed all of these blocks in CMOS 65$nm$ technology where we have used 60$nm$ and 120$nm$ as the minimum length and width of a transistor. All digital gates or components are sized to have minimum area with added buffers to match their drive strength with the required load. All analog components like sense amplifiers, pass-gates etc. are sized accordingly, usually larger than digital ones, to reduce the impact of mismatch and noise. Table \ref{tab:power_overhead} presents the power consumption in terms of overall current in different states of the system. As we have discussed before, XbarPUF can have a larger power consumption due to the state where all the memristors are reset to HRS. Therefore, during state 1 and 2, the power consumption is high but it is fairly small during other times. Thus, the average current of the whole system is roughly 24.80$\mu$A overall with a 0.85V supply voltage. We can further reduce this current by utilizing a dual-voltage scheme, with smaller VDD for digital circuits, and a separate larger VDD for analog and memristive components. 

\begin{table}
\centering
  \caption{Delay overhead of the system in different stages of operation}
  \label{tab:delay_overhead}
  \begin{tabular}{ccl}
     \toprule
     State & Clock cycles & Comment\\
     \midrule
     PUF RESET & 0.5 & 8$\times$\\
     PUF challenge & 0.25 & 8$\times$ short spike\\
     PUF read & 0.25 & 8$\times$ \\
     SRAM write & 4 & overlapped with PUF read\\
     RRAM write & 16+x & for 16 clean bit + `x' bit error\\
     RRAM read & 16 & for 16 bit \\
     Tag generation & 3 & 8 bit Tag \\
  \bottomrule
\end{tabular}
\end{table}

\begin{table*}
\centering
\caption{Performance comparison with state-of-the-art lightweight hardware security techniques}
\label{tab:comparison}
\begin{tabular}{ccccccccl}
     \toprule
     Overhead & \multicolumn{4}{c}{Encryption-Decryption}& \multicolumn{3}{c}{Tag generation} \\ 
      \midrule
       & nanoAES \cite{Mathew_nanoAES:2015} & AES \cite{Amir_AES_hardware:2011} & PRESENT \cite{Bogdanov_PRESET:2007} & This work & Hong \textit{et al.} \cite{Hong_Tag:2012} & Yan \textit{et al.} \cite{Yan_tag:2006} & This work \cite{Majumder_IntegrityCheck:2018} \\
      \midrule
      Avg. Power ($\mu$W) &170&18.5&-&21.08&1920&3923&487\\
      Delay (clock cycles/bit) &2.62&1.75&0.5&0.27&7.5/128&15/128&3/128\\
      Area (NAND G.E.) &2090&2400&1570&856&2339&3835&864\\
  \bottomrule
\end{tabular}
\end{table*}

Table \ref{tab:delay_overhead} shows the delay of our system in terms of required number of clock cycles. We have 4 XbarPUFs in our design and each one is activated twice during either encryption or decryption. Thus response generation from these four XbarPUFs twice would take $m\times$4$\times$2 clock cycles where $m$ is the number of clock cycles required to generate a response from one XbarPUF which is designed to have 1 clock cycle delay ( different phases of same clock cycle for reset, challenge, and read). One row of SRAM are written together and thus our 4$\times$8 SRAM would require 4 clock cycles for a complete write. However, this is overlapped with the response generation phase and thus doesn't add to the overall delay. The slowest phase of our design is when the encrypted data is written one bit at a time to the RRAM. For a `n'-bit key, it would require at least `n' clock cycles and for this prototype system, `n' is 16. In practice, this stage would require more than n clock cycles as `x' number of unreliable or noisy bits would add `x' extra clock cycles. However, this delay can be reduced by allowing to write multiple bits in the RRAM during back-up. Tag generation takes about 3 clock cycles. During decryption, RRAM is read one bit at a time for a total of `n' clock cycles (again n=16 here). The time required during decryption thus would be almost the same as the encryption as the operations are very similar.


We have presented a comparison of resource overhead for our proposed security vs traditional security mechanisms employed in embedded system. Table \ref{tab:comparison} lists the resource overhead for traditional encryption algorithms, nanoAES \cite{Mathew_nanoAES:2015} and PRESENT \cite{Bogdanov_PRESET:2007} and tag generation schemes presented in \cite{Hong_Tag:2012,Yan_tag:2006} alongside our proposed security solution. To calculate the area overhead, we have omitted the circuit blocks that would already be present in a system (e.g. counters, SRAMs, registers etc.). However, the delay and power from these blocks are taken into consideration. Although our proposed system can generate 30 clean bits on average, one can generate a larger key by applying more challenges. This would increase the delay and energy requirement while keeping the overall area and average power the same. Alternatively, larger PUF blocks can be used to generate a larger key with no additional delay at the expense of larger area and power. 

It should be noted that the overhead for existing techniques reported in Table \ref{tab:comparison} are without the overhead associated with generating and storing the key. Thus their actual implementation should have even larger resource overhead. However, this key generation is the main contributing factor to overhead in our proposed system. This comes at the added benefit of random unique keys for each system and the key doesn't need be saved physically. Tag generation results were presented in \cite{Majumder_IntegrityCheck:2018} and shown here again to highlight the lightweight nature compared to other tag generation schemes. 
The overhead associated with this tag generation method is significantly lower than another comparable method as can be seen from Table \ref{tab:comparison}. Overall, the system proposed in this work provides both confidentiality and integrity for the non-volatile back-up memory of a resource-constrained system.

\section{Future Work}
\label{sec:future}
In this work, we have demonstrated the error correction method to generate a secure cryptographic key from the response of an XbarPUF. Then using this PUF based encryption plus integrity checking can be incorporated to an embedded processor to emulate a full system which saves it processor states and other relevant information securely in a non-volatile memory when a power failure occurs. The amount of overhead depending on the frequency of this power failure from a particular energy source can be analyzed. Finally, the overall security gain vs. cost of additional circuitry could be evaluated for a practical device.

For energy harvesting devices where the power profile is relatively predictable, and regular periodic power failure happens, our proposed system can be tweaked to give more accuracy. Suppose for a system based on only solar power would have regular power failure at the end of each day and power would come back on at the beginning of next day. To make up for the temperature difference between these two times (dawn and dusk), the first set of PUF response can be generated at the start of each day. Then during power failure, second set of PUF response can be generated and compared with the first response to have a better estimation of environmental condition and thus more reliable PUF response. PUF key can also be generated beforehand while the system is running, thereby reducing the power and delay requirement during critical back-up situation.

\section{Conclusion}
\label{sec:conc}

We have presented a novel two layer security protocol for a resource constrained IoT during power failure or power saving mode. This is a two layer security where the PUF based OTP encryption being the first layer of protection and the tag based data integrity being the second layer of security. As data forward progression is a common and efficient method for resource limited systems, our proposed scheme adds reasonable security to these systems. We have implemented our whole system in transistor level with emerging memristive technology to get an accurate idea about system implementation. The key generated from a PUF is unclonable and random by nature while being volatile i.e. the key is not saved anywhere and generated only when a back-up operation is needed. Thus we are able to provide unique device-specific security for each different IC even with the same functionality. The target application for our implemented system would be an embedded processor where processor goes to `sleep' or power saving mode and saves state information for a quick wake-up. It can also be useful for batteryless system where power failure could occur and data are backed up in non-volatile memory for forward progression. Our system would fit into any system where aggressive power saving techniques are employed and data is backed-up in an NVM to avoid a large computational penalty. With the increasing number of edge devices with very small resource and little to no security, our proposed idea could be an effective solution to provide practical level of security that these systems can carry.


\section*{Acknowledgment}
The authors would like to thank Abhishek Bhandari and Grayson Bruner of our research group for important discussions on this topic.
\bibliographystyle{IEEEtran}
\bibliography{IEEEabrv,mishuk,mishuk-IOT,garose}

\begin{thebibliography}{10}
\providecommand{\url}[1]{#1}
\csname url@samestyle\endcsname
\providecommand{\newblock}{\relax}
\providecommand{\bibinfo}[2]{#2}
\providecommand{\BIBentrySTDinterwordspacing}{\spaceskip=0pt\relax}
\providecommand{\BIBentryALTinterwordstretchfactor}{4}
\providecommand{\BIBentryALTinterwordspacing}{\spaceskip=\fontdimen2\font plus
\BIBentryALTinterwordstretchfactor\fontdimen3\font minus
  \fontdimen4\font\relax}
\providecommand{\BIBforeignlanguage}[2]{{%
\expandafter\ifx\csname l@#1\endcsname\relax
\typeout{** WARNING: IEEEtran.bst: No hyphenation pattern has been}%
\typeout{** loaded for the language `#1'. Using the pattern for}%
\typeout{** the default language instead.}%
\else
\language=\csname l@#1\endcsname
\fi
#2}}
\providecommand{\BIBdecl}{\relax}
\BIBdecl

\bibitem{IoT-device-number:2015}
\BIBentryALTinterwordspacing
S.~R. Department, ``Internet of things (iot) connected devices installed base
  worldwide from 2015 to 2025 (in billions),'' Nov 2016. [Online]. Available:
  \url{https://www.statista.com/statistics/471264/iot-number-of-connected-devices-worldwide/}
\BIBentrySTDinterwordspacing

\bibitem{Sikder_IoT_privacy_survey:2018}
A.~K. Sikder, G.~Petracca, H.~Aksu, T.~Jaeger, and A.~S. Uluagac, ``A survey on
  sensor-based threats to internet-of-things ({IOT}) devices and
  applications,'' \emph{arXiv preprint arXiv:1802.02041}, 2018.

\bibitem{Acar_IoT_peekaboo:2018}
A.~Acar, H.~Fereidooni, T.~Abera, A.~K. Sikder, M.~Miettinen, H.~Aksu,
  M.~Conti, A.-R. Sadeghi, and A.~S. Uluagac, ``Peek-a-boo: I see your smart
  home activities, even encrypted!'' \emph{arXiv preprint arXiv:1808.02741},
  2018.

\bibitem{Oconnor_IoT_homesnitch:2019}
T.~OConnor, R.~Mohamed, M.~Miettinen, W.~Enck, B.~Reaves, and A.-R. Sadeghi,
  ``Homesnitch: behavior transparency and control for smart home iot devices,''
  in \emph{Proceedings of the 12th Conference on Security and Privacy in
  Wireless and Mobile Networks}, 2019, pp. 128--138.

\bibitem{Babun_IoT_privacy:2019}
L.~Babun, Z.~B. Celik, P.~McDaniel, and A.~S. Uluagac, ``Real-time analysis of
  privacy-(un) aware {IoT} applications,'' \emph{arXiv preprint
  arXiv:1911.10461}, 2019.

\bibitem{Ma:Energy_harvesting_architecture_explore:2015}
K.~Ma, Y.~Zheng, S.~Li, K.~Swaminathan, X.~Li, Y.~Liu, J.~Sampson, Y.~Xie, and
  V.~Narayanan, ``Architecture exploration for ambient energy harvesting
  nonvolatile processors,'' in \emph{2015 IEEE 21st International Symposium on
  High Performance Computer Architecture (HPCA)}.\hskip 1em plus 0.5em minus
  0.4em\relax IEEE, 2015, pp. 526--537.

\bibitem{Manifavas_Lightweight_Crypto:2013}
C.~Manifavas, G.~Hatzivasilis, K.~Fysarakis, and K.~Rantos, ``Lightweight
  cryptography for embedded systems--a comparative analysis,'' in \emph{Data
  Privacy Management and Autonomous Spontaneous Security}.\hskip 1em plus 0.5em
  minus 0.4em\relax Springer, 2013, pp. 333--349.

\bibitem{Amir_AES_hardware:2011}
A.~Moradi, A.~Poschmann, S.~Ling, C.~Paar, and H.~Wang, ``Pushing the limits: A
  very compact and a threshold implementation of {AES},'' in \emph{Advances in
  Cryptology -- EUROCRYPT 2011}, K.~G. Paterson, Ed.\hskip 1em plus 0.5em minus
  0.4em\relax Berlin, Heidelberg: Springer Berlin Heidelberg, 2011, pp. 69--88.

\bibitem{Bogdanov_PRESET:2007}
A.~Bogdanov, L.~R. Knudsen, G.~Leander, C.~Paar, A.~Poschmann, M.~J. Robshaw,
  Y.~Seurin, and C.~Vikkelsoe, ``Present: An ultra-lightweight block cipher,''
  in \emph{International workshop on cryptographic hardware and embedded
  systems}.\hskip 1em plus 0.5em minus 0.4em\relax Springer, 2007, pp.
  450--466.

\bibitem{Suh:2005}
G.~E. Suh, C.~W. O'Donnell, I.~Sachdev, and S.~Devadas, ``Design and
  implementation of the {AEGIS} single-chip secure processor using physical
  random functions,'' in \emph{Proc. of the 32nd Annual Int. Symp. on Comput.
  Architecture}, 2005, pp. 25--36.

\bibitem{Strukov:2008}
D.~B. Strukov, G.~S. Snider, D.~R. Stewart, and R.~S. Williams, ``The missing
  memristor found,'' \emph{Nature}, vol. 453, pp. 80--83, May 2008.

\bibitem{Mesbah-JETC:2017}
M.~Uddin, M.~B. Majumder, K.~Beckmann, H.~Manem, Z.~Alamgir, N.~C. Cady, and
  G.~S. Rose, ``Design considerations for memristive crossbar physical
  unclonable functions,'' \emph{J. Emerg. Technol. Comput. Syst.}, vol.~14,
  no.~1, pp. 2:1--2:23, sep 2017.

\bibitem{Majumder_IntegrityCheck:2018}
M.~B. Majumder, M.~S. Hasan, M.~Uddin, and G.~S. Rose, ``A secure integrity
  checking system for nanoelectronic resistive ram,'' \emph{IEEE Transactions
  on Very Large Scale Integration (VLSI) Systems}, pp. 1--14, 2018.

\bibitem{Mesbah_TRNG:2019}
M.~Uddin, M.~S. Hasan, and G.~S. Rose, ``On the theoretical analysis of
  memristor based true random number generator,'' in \emph{Proceedings of the
  2019 on Great Lakes Symposium on VLSI}.\hskip 1em plus 0.5em minus
  0.4em\relax ACM, 2019, pp. 21--26.

\bibitem{TaOx_Yang:2010}
J.~J. Yang, M.~Zhang, J.~P. Strachan, F.~Miao, M.~D. Pickett, R.~D. Kelley,
  G.~Medeiros-Ribeiro, and R.~S. Williams, ``High switching endurance in {TaOx}
  memristive devices,'' \emph{Applied Physics Letters}, vol.~97, no.~23, p.
  232102, 2010.

\bibitem{TiO_Medeiros:2011}
G.~Medeiros-Ribeiro, F.~Perner, R.~Carter, H.~Abdalla, M.~D. Pickett, and R.~S.
  Williams, ``Lognormal switching times for titanium dioxide bipolar
  memristors: origin and resolution,'' \emph{Nanotechnology}, vol.~22, no.~9,
  p. 095702, 2011.

\bibitem{Rose:2013a}
G.~S. Rose, N.~McDonald, L.~Yan, B.~Wysocki, and K.~Xu, ``Foundations of
  memristor based {PUF} architectures,'' in \emph{Proc. of the {IEEE/ACM} Int.
  Symp. on Nanoscale Architectures ({NANOARCH})}, July 2013, pp. 52--57.

\bibitem{Jiang:2017_Memristor_TRNG}
H.~Jiang, D.~Belkin, S.~E. Savel\'ev, S.~Lin, Z.~Wang, Y.~Li, S.~Joshi,
  R.~Midya, C.~Li, M.~Rao \emph{et~al.}, ``A novel true random number generator
  based on a stochastic diffusive memristor,'' \emph{Nature communications},
  vol.~8, no.~1, p. 882, 2017.

\bibitem{Hamdioui:2015_memristor_in_memory_compute}
S.~Hamdioui, L.~Xie, H.~A.~D. Nguyen, M.~Taouil, K.~Bertels, H.~Corporaal,
  H.~Jiao, F.~Catthoor, D.~Wouters, L.~Eike \emph{et~al.}, ``Memristor based
  computation-in-memory architecture for data-intensive applications,'' in
  \emph{Proceedings of the 2015 design, automation \& test in Europe conference
  \& exhibition}.\hskip 1em plus 0.5em minus 0.4em\relax EDA Consortium, 2015,
  pp. 1718--1725.

\bibitem{Xu_RRAM_memory:2011}
C.~{Xu}, X.~{Dong}, N.~P. {Jouppi}, and Y.~{Xie}, ``Design implications of
  memristor-based rram cross-point structures,'' in \emph{2011 Design,
  Automation Test in Europe}, March 2011, pp. 1--6.

\bibitem{Zidan_memristor_future:2018}
M.~A. Zidan, J.~P. Strachan, and W.~D. Lu, ``The future of electronics based on
  memristive systems,'' \emph{Nature Electronics}, vol.~1, no.~1, p.~22, 2018.

\bibitem{Mesbah_SA:2018}
M.~{Uddin} and G.~S. {Rose}, ``A practical sense amplifier design for
  memristive crossbar circuits (puf),'' in \emph{2018 31st IEEE International
  System-on-Chip Conference (SOCC)}, Sep. 2018, pp. 209--214.

\bibitem{Suh:2007}
G.~Suh and S.~Devadas, ``Physical unclonable functions for device
  authentication and secret key generation,'' in \emph{44th ACM/EDAC/IEEE
  Design Automation Conference (DAC)}, June 2007, pp. 9--14.

\bibitem{paral2011reliable}
Z.~Paral and S.~Devadas, ``Reliable and efficient {PUF}-based key generation
  using pattern matching,'' in \emph{IEEE Int. Symp. on Hardware-Oriented
  Security and Trust (HOST)}.\hskip 1em plus 0.5em minus 0.4em\relax IEEE,
  2011, pp. 128--133.

\bibitem{Rose:2015}
G.~Rose and C.~Meade, ``Performance analysis of a memristive crossbar {PUF}
  design,'' in \emph{52nd ACM/EDAC/IEEE Design Automation Conference (DAC)},
  June 2015, pp. 1--6.

\bibitem{Kaisheng_energy_harvest_NVM:2015}
K.~Ma, X.~Li, S.~Li, Y.~Liu, J.~J. Sampson, Y.~Xie, and V.~Narayanan,
  ``Nonvolatile processor architecture exploration for energy-harvesting
  applications,'' \emph{IEEE Micro}, vol.~35, no.~5, pp. 32--40, 2015.

\bibitem{Roarke_OTP:2013}
R.~Horstmeyer, I.~M. Vellekoop, S.~Assawaworrarit, B.~Judkewitz, and C.~Yang,
  ``Physical key-protected one-time pad,'' \emph{Scientific Reports, Nature},
  vol.~3, no. 3543, December 2013.

\bibitem{Mesbah:2016}
M.~Uddin, M.~B. Majumder, G.~S. Rose, K.~Beckmann, H.~Manem, Z.~Alamgir, and
  N.~C. Cady, ``Techniques for improved reliability in memristive crossbar
  {PUF} circuits,'' in \emph{IEEE Comp. Society Annual Symp. on VLSI (ISVLSI)},
  July 2016, pp. 212--217.

\bibitem{Mesbah_CEM:2018}
M.~{Uddin}, B.~{Majumder}, and G.~S. {Rose}, ``Nanoelectronic security designs
  for resource-constrained internet of things devices: Finding security
  solutions with nanoelectronic hardwares,'' \emph{IEEE Consumer Electronics
  Magazine}, vol.~7, no.~6, pp. 15--22, Nov 2018.

\bibitem{Mesbah_TNANO:2017}
M.~Uddin, M.~B. Majumder, and G.~S. Rose, ``Robustness analysis of a memristive
  crossbar {PUF} against modeling attacks,'' \emph{IEEE Transactions on
  Nanotechnology}, vol.~16, no.~3, pp. 396--405, May 2017.

\bibitem{Ruhrmair:2013}
U.~Ruhrmair, J.~Solter, F.~Sehnke, and X.~Xu, ``{PUF} modeling attacks on
  simulated and silicon data,'' \emph{{IEEE} Trans. on Inform. Forensics and
  Security}, vol.~8, pp. 1876--1891, August 2013.

\bibitem{Maiti:2013}
A.~Maiti, V.~Gunreddy, and P.~Schaumont, ``\BIBforeignlanguage{English}{A
  systematic method to evaluate and compare the performance of physical
  unclonable functions},'' in \emph{\BIBforeignlanguage{English}{Embedded
  Systems Design with FPGAs}}, P.~Athanas, D.~Pnevmatikatos, and N.~Sklavos,
  Eds.\hskip 1em plus 0.5em minus 0.4em\relax Springer New York, 2013, pp.
  245--267.

\bibitem{OTP_wiki:2020}
\BIBentryALTinterwordspacing
Wikipedia, ``One-time pad,'' Feb 2020. [Online]. Available:
  \url{https://en.wikipedia.org/wiki/One-time_pad}
\BIBentrySTDinterwordspacing

\bibitem{Hospodar-modeling:2012}
G.~Hospodar, R.~Maes, and I.~Verbauwhede, ``Machine learning attacks on 65nm
  arbiter {PUF}s: Accurate modeling poses strict bounds on usability,'' in
  \emph{2012 IEEE International Workshop on Information Forensics and Security
  (WIFS)}, Dec 2012, pp. 37--42.

\bibitem{scikit-learn}
F.~Pedregosa, G.~Varoquaux, A.~Gramfort, V.~Michel, B.~Thirion, O.~Grisel,
  M.~Blondel, P.~Prettenhofer, R.~Weiss, V.~Dubourg, J.~Vanderplas, A.~Passos,
  D.~Cournapeau, M.~Brucher, M.~Perrot, and E.~Duchesnay, ``Scikit-learn:
  Machine learning in {P}ython,'' \emph{Journal of Machine Learning Research},
  vol.~12, pp. 2825--2830, 2011.

\bibitem{Kobayashi:1993}
T.~Kobayashi, K.~Nogami, T.~Shirotori, and Y.~Fujimoto, ``A current-controlled
  latch sense amplifier and a static power-saving input buffer for low-power
  architecture,'' \emph{IEEE J. of Solid-State Circuits}, vol.~28, no.~4, pp.
  523--527, Apr 1993.

\bibitem{Mathew_nanoAES:2015}
S.~Mathew, S.~Satpathy, V.~Suresh, M.~Anders, H.~Kaul, A.~Agarwal, S.~Hsu,
  G.~Chen, and R.~Krishnamurthy, ``340 mv--1.1 v, 289 gbps/w, 2090-gate
  {NanoAES} hardware accelerator with area-optimized encrypt/decrypt gf (2 4) 2
  polynomials in 22 nm tri-gate cmos,'' \emph{IEEE Journal of Solid-State
  Circuits}, vol.~50, no.~4, pp. 1048--1058, 2015.

\bibitem{Hong_Tag:2012}
M.~Hong, H.~Guo, and S.~X. Hu, ``A cost-effective tag design for memory data
  authentication in embedded systems,'' in \emph{Proceedings of the 2012
  international conference on Compilers, architectures and synthesis for
  embedded systems}.\hskip 1em plus 0.5em minus 0.4em\relax ACM, 2012, pp.
  17--26.

\bibitem{Yan_tag:2006}
C.~Yan, D.~Englender, M.~Prvulovic, B.~Rogers, and Y.~Solihin, ``Improving
  cost, performance, and security of memory encryption and authentication,'' in
  \emph{ACM SIGARCH Computer Architecture News}, vol.~34, no.~2.\hskip 1em plus
  0.5em minus 0.4em\relax IEEE Computer Society, 2006, pp. 179--190.

\end{thebibliography}
\end{document}